\documentstyle[psfig]{l-aa}

\def\be{$^9$Be}
\def\ab{($^9$Be/H)}

\def\lis{$^6$Li}

\def \zp{$\zeta$~Per}
\def \res{$\lambda/\Delta\lambda$}
\def \ki2{$\chi^2$}

\def\ga{\mathrel{\mathchoice {\vcenter{\offinterlineskip\halign{\hfil
$\displaystyle##$\hfil\cr>\cr\sim\cr}}}
{\vcenter{\offinterlineskip\halign{\hfil$\textstyle##$\hfil\cr>\cr\sim\cr}}}
{\vcenter{\offinterlineskip\halign{\hfil$\scriptstyle##$\hfil\cr>\cr\sim\cr}}}
{\vcenter{\offinterlineskip\halign{\hfil$\scriptscriptstyle##$\hfil
\cr>\cr\sim\cr}}}}}
\def\la{\mathrel{\mathchoice {\vcenter{\offinterlineskip\halign{\hfil
$\displaystyle##$\hfil\cr<\cr\sim\cr}}}
{\vcenter{\offinterlineskip\halign{\hfil$\textstyle##$\hfil\cr<\cr\sim\cr}}}
{\vcenter{\offinterlineskip\halign{\hfil$\scriptstyle##$\hfil\cr<\cr\sim\cr}}}
{\vcenter{\offinterlineskip\halign{\hfil$\scriptscriptstyle##$\hfil
\cr<\cr\sim\cr}}}}}
\begin{document}

\thesaurus{08.09.2; 09.01.1}
\title{A very reduced upper limit on the interstellar abundance
of~ ~beryllium.\thanks{Based on observations collected at the Canada-France-Hawaii
Telescope, Hawaii, USA.}}
\author{Guillaume H\'ebrard \inst{1}
	\and Martin Lemoine \inst{2}
	\and Roger Ferlet \inst{1}
	\and Alfred Vidal--Madjar \inst{1}}

\offprints{G. H\'ebrard (hebrard@iap.fr)}
\institute{Institut d'Astrophysique de Paris, CNRS, 98 bis boulevard Arago,
75014 Paris, France
	\and Department of Astronomy \& Astrophysics, Enrico Fermi Institute,
The University of Chicago, Chicago IL60637-1433, USA}
\date{Received January 22, accepted February 25, 1997}
\maketitle
\begin{abstract}
We present the results of observations of the $\lambda3130.4$\AA \ 
interstellar absorption line of
\be $\,\sc ii$ in the direction of \zp. The data were obtained at the
Canada-France-Hawaii 3.6m Telescope using the Coud\'e f/4 Gecko
spectrograph at a resolving power \res$\,\simeq1.1\times10^5$, and a
signal-to-noise ratio ${\rm S/N}\simeq2000$. The \be
$\,\sc ii$ line is not detected, and we obtain an upper limit on the equivalent width
$W_{3130.4}\leq30$ $\mu$\AA. This upper limit is 
7 times below the lowest upper limit ever reported hitherto. The
derived interstellar abundance
is \ab$\,\leq7\times10^{-13}$, not corrected for the depletion of \be~onto
interstellar grains; it corresponds to an upper limit $\delta_{Be}\leq-1.5$
dex on the depletion factor of \be.
As such, it argues in favour of models of formation
of dust grains in stellar atmospheres.
\end{abstract}

\section{Introduction}
Beryllium is created in Big Bang nucleosynthesis (BBN) with an extremely low
primordial abundance, \ab$_p<10^{-14}$. Subsequently, it is
solely formed in spallation reactions of galactic cosmic rays (GCR)
interacting with interstellar C, N, O atoms, and is thoroughly destroyed
through astration of interstellar gas. This simple scenario allows to account for the observed
Pop I abundance of \be, \ab$_{PopI}\simeq1.3\times10^{-11}$ (Boesgaard
1976), the solar
abundance \ab$_{\odot}\simeq1.4\times10^{-11}$ (Chmielewski et al. 1975) and
the meteoritic abundance
\ab$_{met}\simeq2.6\times10^{-11}$ (Anders \& Grevesse 1989). For this reason, \be~together with \lis,
which shares a similar evolutionary picture, are used as tracers of cosmic ray
spallation~activity.

For our present purpose, the main importance of the interstellar
abundance of \be~is related to the physics of formation of dust grains, through
the depletion factor of \be , $\delta_{Be}$ (Snow et al. 1979). Field (1974)
observed a correlation between the underabundance of an element in the
interstellar medium (ISM) and the condensation temperature of that element,
defined as the threshold at which half of the gaseous phase has gone to solid
state; this suggests that dust grains were formed under equilibrium
pressure in late-type giants atmospheres or stellar nebulae. On the other hand,
Snow (1975), noticing a similar correlation trend between the depletion factors
of chemical elements and their first ionization potentials (except for three
elements), suggested that dust
grains were formed by collisions in the ISM. In this case, depletion should
increase with time, elements should be selectively depleted
({\it i.e.}~the element-to-element depletion ratio should vary from one line of
 sight to another and the amount of
depletion should depend on the cloud density. Similar conclusions have been
subsequently reached by Barlow (1978), and Duley \& Millar (1978). This
situation is not settled as yet, since many more observations have revealed
that:
(i) depletions effectively vary from sightline to sightline; (ii) but there is
no evidence of element-to-element depletions ratio variation (Joseph 1988, and
references therein). As well, it has been shown that the condensation temperature may
not always be a good indicator of depletion, most notably with respect to the
phosphorus/iron ratio of depletion (Jura \& York 1978). Note that 
in the above presentation and all throughout the paper, 
we only refer to rather diffuse clouds,
and not to denser clouds. In effect, depletion within cloud discussed here
could be quite different from the denser ones in which the material is
probably protected from grain-disrupting shocks and the density
high enough to initiate species-specific depletion~mechanisms.

Beryllium presents the advantage of being a clear element discriminator of
these various approaches. Its depletion factor is predicted to be
 $\simeq-0.2$ dex
in a correlation with first ionization potential, and $\simeq-1.5$ dex in a
correlation with condensation temperature (Snow et al. 1979; Boesgaard 1985).
However, \be \ has never been detected in the ISM so that, its actual
depletion factor is largely~unknown.

Here, we report on our observations of the $\lambda3130.4$\AA $\,$ interstellar
absorption line of \be $\,\sc ii$ in the direction of~\zp. Previous
unsuccessful attempts are discussed in Sec.2, together with our observations;
their analysis are discussed in Sec.3, and their results in Sec.4.

\section{Observations.}
No firm detection of \be~has ever been reported in the ISM because of the
extreme difficulty of this observation. \be $\,\sc ii$, the dominant ionization
stage of \be~in the ISM (the first ionization potential of \be \ is 9.3 eV, and
the second 18.2 eV), can only be observed through the resonance doublet at
$\lambda3130.420/3131.066$\AA, {\it i.e.} in the near--ultraviolet where the
atmospheric absorption is important. On the basis of the cosmic abundance of
\be , \ab$_{cosm}\simeq\,$\ab$_{met}\simeq2.6\times10^{-11}$
[\ab$_{cosm}\neq\,$\ab$_{\odot}$ cause of the depletion of \be \ by nuclear
reactions at the bottom of the convection zone of the Sun (Anders \& Grevesse
1989)], the hydrogen column density toward \zp
, $N$(H)=1.6$\times10^{21}$ cm$^{-2}$ (Savage et al. 1977), and a depletion
 factor $\delta_{Be}$ between $-0.1$ and $-2.0$ dex, 
 one should expect a column density
$N$(Be$\,\sc ii$) ranging from $3\times10^{10}$ to $4\times10^8$ cm$^{-2}$,
hence an equivalent width $W_{3130.4}$ from 900 to $10\,\mu$\AA . In this
optically thin case, we obtain the column density $N$ (in cm$^{-2}$) from the
equivalent width $W$ (in \AA ) by the expression:
$$N=1.13\times10^{20}\frac{W}{\lambda^2f},$$
where $f$ is the oscillator strength and $\lambda$ the wavelength (in
\AA). The upper limits to the equivalent widths
of interstellar \be \ that have been reported up to now are listed in Table~1.

\begin{center}
\begin{table*}
\caption[]{Previous attempts at detecting interstellar \be~and upper limits
derived. These abundances are not corrected for the depletion of \be~onto
dust grains.}
\topskip 5cm
\begin{flushleft}
\begin{tabular*}{18.5cm}{p{8cm}ccccccc}
\hline
Authors & Targets & \ \ & \ \ & Equivalent widths & \ \ & \ \ & Abundances \bigskip\\
\hline
Herbig (1968) & $\zeta$ Oph & \ \ & \ \ & $<2.5$ m\AA & \ \ & \ \ & $<1.4\times10^{-11}$\\
Boesgaard (1974) & 22 stars & \ \ & \ \ &  & \ \ & \ \ & $<5\times10^{-11}$ \\
Chaffee \& Lutz (1977) & $\zeta$ Per & \ \ & \ \ & $<0.6$ m\AA & \ \ & \ \ & $<1.3\times10^{-11}$\\
York, Meneguzzi \& Snow (1982) & $\zeta$ Oph & \ \ & \ \ & $<1$ m\AA & \ \ & \ \ & $<2\times10^{-11}$\\
York, Meneguzzi \& Snow (1982) & $\sigma$ Sco & \ \ & \ \ & $<1$ m\AA & \ \ & \ \ & 
$<1.3\times10^{-11}$\\
Boesgaard (1985) & $\zeta$ Per & \ \ & \ \ & $<0.23$ m\AA & \ \ & \ \ & $<4.8\times10^{-12}$ \\
Boesgaard (1985) & $\delta$ Sco & \ \ & \ \ & $<0.36$ m\AA & \ \ & \ \ & $<8.4\times10^{-12}$ \\
Baade \& Crane (1991) & $\zeta$ Oph & \ \ & \ \ & $<0.3$ m\AA & \ \ & \ \ & $<6\times10^{-12}$ \\
\hline
This work & $\zeta$ Per & \ \ & \ \ & $<0.03$ m\AA & \ \ & \ \ & $<7\times10^{-13}$ \\
\hline
\end{tabular*}
\end{flushleft}
\end{table*}
\end{center}

Our observations were conducted in January 1994 and October 1995 at the
Canada--France--Hawaii Telescope, whose altitude (4200 m) allows for a good
UV transparency. We used the spectrograph
Coud\'e f/4 Gecko at high resolving power $R\simeq1.1\times10^5$ 
(equivalently 0.029~\AA, 2.8~km.s$^{-1}$, or 2.8~pixels).
This spectrograph carries a Richardson image slicer. The detector is a 2K CCD
with a pixel size of 15 $\mu$m. Its quantum efficiency is about 0.7 at 3100
\AA. We took care during the observations to shift the central wavelength from
night to night in order to identify eventual systematic features on the CCD
detector. As well, thorium-argon
 lamp calibration exposures were recorded each night, interspaced with
stellar exposures, in order to achieve as high a level of wavelenght accuracy
as possible. This is necessary in that one goes blind looking for a line that
may appear only after all individual spectra have been correctly shifted and
properly averaged. Over the six observing nights, calibration root
mean square range between 1 and 2 m\AA \ ({\it i.e.} between 1/10$^{th}$ and
1/5$^{th}$ of a pixel).
Flat fields were performed on a platinum lamp at the
beginning and the end of night in order to obtain a very high level flat
field while optimizing the integration time spent on \zp.

\section{Data reduction and analysis.}
The data were reduced using the IRAF and MIDAS softwares. It was found that
scattered
light is present in the spectrograph at a non-negligible level. Moreover, due
to the presence of a cross-disperser (which projects different spectra from
different orders perpendicularly to the dispersion) and to the small
inter-order spacing in the near-UV, there resulted a slight overlap
of the scattered light with observing adjacent orders at the time of the first
run. This overlap was not present during the second run because of a
better adjusting of the spectrograph.
Even when there is an overlap, the scattered light could nonetheless
be interpolated, and the remaining background level after
removal did not exceed $\simeq1\%$ in all CCD frames, so that the zero
flux level should be precisely known.

 It proved difficult to flat-field the
spectra because of the image slicer device. In effect, the 
5 slices of signal are here reconstructed one on top of another on the
CCD, instead of being spread out perpendicularly to the dispersion. 
We noticed that the signal on the CCD was not uniform, meaning
that probably the reconstruction is not perfect. However, this non-uniformity
of the signal might also be due to a non-perfect focussing of the four
separate grating of the mosaic. As a result, a non-negligible part of
the image on the CCD typically had to be dropped off before flat-fielding and
averaging to a single spectrum, in order to preserve a precise flat-fielding.
Before co-adding the different spectra (about fifty of typically 30-45 min
integration time each), each spectrum was corrected to the heliocentric rest
frame. Different statistical filters for co-addition were applied to
different set of spectra, and at the end, the average spectrum showing
the highest signal-to-noise ratio in the vicinity of
$\lambda3130$\AA~was kept.

The aspect of the average normalized spectrum of \zp \ in the
$\lambda3130$\AA \ region is shown in Fig.1. An enlargement of the final
spectrum around the \be $\,\sc ii$ line is shown in Fig.2 where $1\sigma$ 
error bars are also plotted.
 It is the weighted average of 44 individual spectra, for a
total integration time of 25 h. We reach a signal-to-noise ratio of
$\simeq2000$ per pixel in the vicinity of the expected line 
(${\rm S/N}\simeq3000$ per resolution element).

\begin{figure}      
\psfig{figure=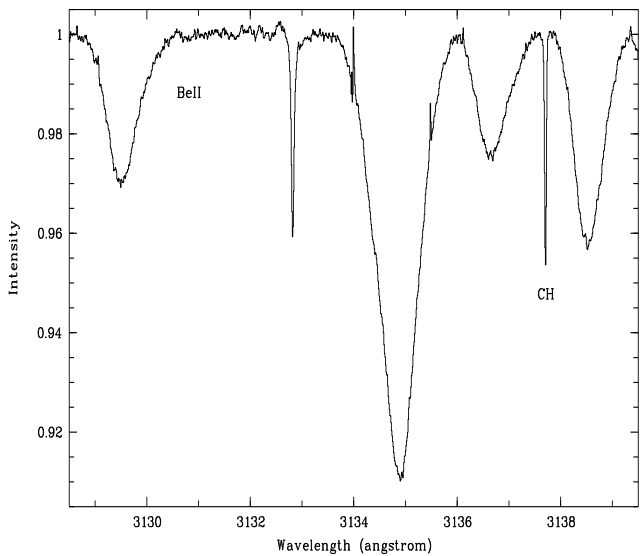,angle=270,height=8cm,width=9cm}
\caption[]{Spectrum of $\zeta$ Per. The resolving
power is $\lambda/\Delta\lambda\simeq1.1\times10^5$.
The \be $\,\sc ii$ doublet is expected at 3130.6~\AA \ and 3131.2~\AA.}
\end{figure}

\begin{figure}
\psfig{figure=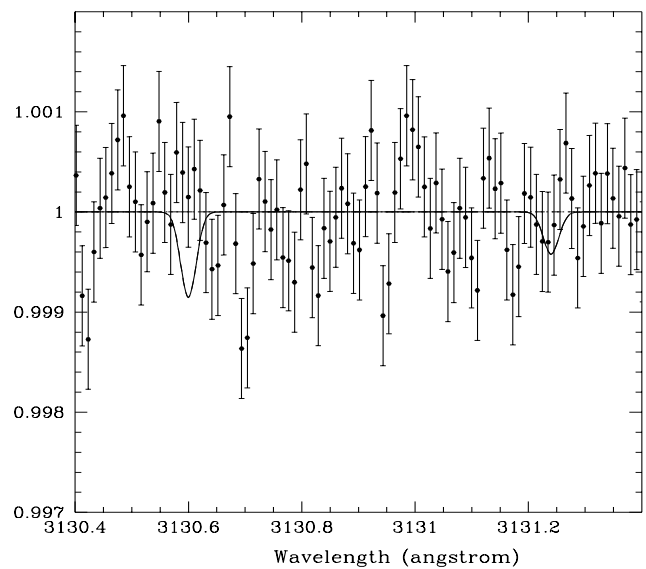,height=8.4cm,width=9.5cm}
\caption[]{Final spectrum of \zp \ enlarged where the the \be \ doublet is
expected. Solid line corresponds to the limiting 30~$\mu$\AA \ detectable 
equivalent width at 3$\sigma$ (see text).}
\end{figure}

Only one absorption line in the spectrum of Fig.1 is identified: the CH line
at 3137.53 \AA \ (oscillator strength $f=1.2\times10^{-3}$, Chaffee \& Lutz 
1977) measured at $3137.706\pm0.006$ \AA, {\it i.e.} at a heliocentric
radial velocity of $16.8\pm0.6$ km.s$^{-1}$.
The Gaussian full width at half maximum (FWHM) for this line is $38\pm8$
m\AA, and the equivalent width $2.0\pm0.2$ m\AA , corresponding to a column
density of $1.9\pm0.2\,\times10^{13}\;{\rm cm}^{-2}$ (optically thin line).
During the first run another CH line at 3143.15 \AA \ was observed with an
equivalent width of $5.9\pm0.6$ m\AA \ corresponding exactly to the $f$ ratio
between the two CH lines.

The widths of the 4 broad spectral features are~$\ga65$~km.s$^{-1}$,
 in agreement with the
stellar rotational~velocity $v.sini=60$ km.s$^{-1}$, confirming their
photospheric origin. The absorption line near 3133 \AA \
is puzzling. Since its position has shifted by $\simeq0.4$~\AA \ 
({\it i.e.}~$\simeq~40$~km.s$^{-1}$) between the two observing 
runs, its origin could be instrumental.

\

The radial velocity of the interstellar CH line found here at $\simeq16.8$
km.s$^{-1}$ differs slightly from previous measurements with different atomic
or ionic lines 
at 14~km.s$^{-1}$ (see {\it e.g.} Hobbs 1978 or Welty et al. 1994 and~1996).
Nevertheless, the difference is within the instrumental~resolutions.

We have searched for the \be$\,\sc ii$ lines at these two velocities.
Whatever the velocity is, the detection is not
convincing. However, the absence of detection at such a high
signal-to-noise ratio and resolution translates into a very reduced upper
limit on the beryllium column density.

To obtain this upper limit, we considered that the strongest \be \ line at
3130.420 \AA \ has the same
FWHM as the CH line. We assume that it has a depth $\la 0.1$\% (the standard
 deviation of the
pixel's value in this spectral
region is $\simeq\pm 0.05\%$). The Gaussian gives then an equivalent width of
about 30 $\mu$\AA . In effect, the limiting detectable equivalent
width $W_{lim}$ at $3\sigma$ is given by 
$W_{lim}\equiv{3\Delta\lambda\over S/N}$. With our S/N per resolution element
$\Delta\lambda$ of $\simeq 3000$, we obtain $W_{lim}\simeq30\,\mu$\AA.

We can then deduce an upper limit on the column~density. In this optically
thin case, we obtain 
$N(^9{\rm Be}\,\sc ii)\simeq1.0\times10^9\;{\rm cm}^{-2}$
($f=0.3382$ for this line, Morton 1991).
The spectrocopic data related to these two lines are known from theoretical
calculations. These lines were already detected in stars revealing no
discrepancies with respect to their $f$-values. The absorptions corresponding 
to this column density are shown as solid line in Fig.2 at a radial velocity 
of 16.8~km.s$^{-1}$ (the result is similar for a radial velocity of
14~km.s$^{-1}$).

We assume now that at least 90\% of the interstellar beryllium is present in the
first ionization stage \be $\,\sc ii$ (Boesgaard 1985). This is supported by
the ratios between ionization stages of others elements. For example, in this
same line of sight, $N({\rm Mg}\,\sc i)$/$N({\rm Mg}\,\sc ii)\leq10^{-2}$ and
$N({\rm S}\,\sc iii)$/$N({\rm S}\,\sc ii)\leq10^{-3}$ (Snow 1977).
Taking the hydrogen column density toward \zp \ 
$N({\rm H})\,=\,1.6\times10^{21}\;{\rm cm}^{-2}$ (Savage et al. 1977), 
we thus deduce an
upper limit of the interstellar abundance for $^9$Be toward $\zeta$ Per:
$$ (^9{\rm Be}/{\rm H})_{\zeta\,Per}\leq7\times10^{-13}\;.$$
This abundance is not corrected for \be~depletion onto dust grains.

\section{Discussion and conclusions.}
Our interstellar abundance of \be~is at least 35 times less
than the cosmic abundance, \ab$_{cosm}\simeq2.6\times10^{-11}$. It
corresponds to a depletion factor $\delta_{Be}\leq-1.5$ dex. This is a new and
much more stringent upper limit compared to previous ones
($\delta_{Be}\leq-0.4$, Boesgaard 1985).

As explained in Sec.1, our present upper limit largely favours the
Field (1974) model of dust grain formation in stellar material. In effect,
the predicted depletion for the condensation temperature of \be \ 
($\simeq1250$~K) is $\simeq-1.5$~dex while the Snow (1975) model of dust grain formation by
chemical trapping predicts $\delta_{Be}\simeq-0.2$~dex.

As already said, the observed absence of selective depletions among chemical 
elements on different sightlines (Joseph 1988) argues also
against the model of Snow (1975).

However, the condensation
temperature curve might not always be a good indicator of depletion, since, for
instance, P is always ten times less depleted than Fe although they have the
same condensation temperature (Jura \& York 1978). This may
be reconciled with the model of Field (1974), if one takes into account the blocking
of P depletion in stellar atmospheres through the formation of stable
molecules ({\it e.g.} PN), which would later dissociate in the ISM (Gail \&
Sedlmayr 1986). Still, Joseph (1988) has shown evidence of a physical process
acting on grains in the ISM, since the overall level of depletion is found to
vary from line of sight to line of sight.
Joseph (1988) proposes that dust grains are indeed formed in a first stage
in stellar material, where the element-to-element depletion ratios are
reproduced; in a second stage, grain destruction in passing shock fronts
would account for the overall variation of depletion. In order to preserve the
observed constancy of the element-to-element depletion ratio, Joseph (1988)
argues that the most depleted elements are locked in grain cores, mainly
Fe, Si, Ca, and the less depleted are trapped in the mantles, mainly P, Mg, S.
These nonvolatile mantles could form in the aftermaths of shocks, and hence
protect the grain core from~destruction.

Beryllium fits in this scenario: \be$\,\sc ii$ as Mg$\,\sc ii$ is one valence, Be,
Mg and P have similar condensation temperatures, and furthermore, Be and P have
similar first ionization potentials. One should therefore expect to find
beryllium trapped in the grain mantles, and to observe a strong correlation of the
depletion factors of these elements on various lines of sight. Our upper limit
$\delta_{Be}\leq-1.5$ is indeed in agreement with the depletion factor
$\delta_{Mg}\simeq-1.3$~dex (Boesgaard 1985)
on the same line of sight. Its discrepancy with the
value $\delta_P\simeq-0.7$~dex (Boesgaard 1985)
could then be associated with the abnormal P/Fe
depletion ratio, {\it i.e.} with the undepletion of phosphorus in stellar
atmospheres (see above). We can hope that a future detection of interstellar
beryllium will not infirm this scenario since apparently we should not be far
from an actual detection.

\end{document}